-------------------------------------------------------------------

\documentstyle[preprint,aps]{revtex}
\newcommand {\be}{\begin{equation}}
\newcommand {\ee}{\end{equation}}

\narrowtext

\begin{document}
\draft

\title{Chronotopic Lyapunov Analysis: \\
(I) a Comprehensive Characterization of 1D Systems}
\author{Stefano Lepri\\
{\it Dipartimento di Fisica, Universit\'a di Bologna and
Istituto Nazionale di Fisica Nucleare\\
I-40127 Bologna, Italy} \\[0.2 cm]
Antonio Politi\\
{\it Istituto Nazionale di Ottica and
Istituto Nazionale di Fisica Nucleare\\
I-50125 Firenze, Italy}\\[0.2 cm]
Alessandro  Torcini\\
{\it Theoretische Physik, Bergische Universit\"at-Gesamthochshule Wuppertal\\
D-42097 Wuppertal, Germany}\\}

\date{\today}

\maketitle
\begin{abstract}
Instabilities in 1D spatially extended systems are studied with the
aid of both temporal and spatial Lyapunov exponents.
A suitable representation of the spectra allows a compact description
of all the possible disturbances in tangent space.
The analysis is carried out for chaotic and periodic spatiotemporal
patterns. Singularities of the spectra and localization properties
of the associated Lyapunov vectors are discussed.
\end{abstract}

\narrowtext
\vskip 1 truecm


\section{Introduction}

Linear stability analysis of chaotic dynamics is usually concerned with the
problem of measuring the divergence of nearby trajectories \cite{eck}.
In extended systems, the spatial dependence of the state variable requires
considering also propagation phenomena and, more generally, spatial
inhomogeneities \cite{mannev}.
The full characterization of a generic perturbation involves
both its temporal and spatial growth rates as complementary measures of
its instability properties. Two classes of Lyapunov exponents have been
separately introduced for this purpose: the former aims at describing
the temporal evolution of disturbances with an exponential profile in
space \cite{Torcini}; the latter deals with the spatial shape
of a perturbation defined on a given site at all times \cite{localiz}.
These, which will be called as {\it temporal} and {\it spatial} exponents,
represent the starting point of this paper. Their definitions are reported
in section III.

Other indicators have been introduced in the attempt of characterizing the
mechanisms of information transport. In this case, the object of study is the
propagation of initially localized disturbances and it led to the introduction
of the comoving Lyapunov exponents \cite{deissler,pikocha}.
Moreover, the Lyapunov analysis of a given (1+1)D pattern can, in
principle, be carried out along any direction. In fact, by performing a
spatio-temporal rotation, i.e. by combining the role of space and time,
it is possible to define still another class of ``rotated'' exponents
\cite{bidime}.\par

It has been already shown that the maximum comoving Lyapunov exponent can
be obtained from the maximum temporal exponent through a Legendre transform
\cite{Torcini}. Moreover, the Kolmogorov-Sinai entropy turns out to be
independent of the propagation direction along a spatio-temporal pattern
\cite{bidime}. Therefore, there must exist strong relationships among the
several classes of exponents so far introduced. A particularly appealing
problem is that of identifying the independent indicators which are necessary
and sufficient for a complete characterization of a linear instability.

The present work, divided into two parts, intends to represent a first step
towards a general and coherent theory. In this first part, we start from
the observation that the most general perturbation is identified by two
rates $\lambda$, $\mu$, describing its growth in time and space, respectively.
The number of linearly independent perturbations can be expressed by two
integrated densities, $n_\lambda(\mu,\lambda)$ and $n_\mu(\mu,\lambda)$. They
provide a dual representation of the problem since any perturbation is
unambiguously identified by the pair $(n_\mu,n_\lambda)$ as well as by
$(\mu,\lambda)$. The properties of these indicators are studied in several
models of coupled map lattices (CML), since we are confident that the tools
and the results apply equally well to systems with continuous space and time
variables.
The relationship with the indicators arising from reference-frame-dependent
analysis of disturbances will be thoroughly discussed in the
second part \cite{lyap2}.

In section II, several classes of CML-models are recalled, which have
been designed to mimick reaction-diffusion systems, many-degrees-of-freedom
Hamiltonian dynamics, and systems with a conserved order parameter.
As already anticipated, in section III spatial and temporal spectra are
formally defined.
In section IV, we introduce the so-called $(\mu,\lambda)$-plane, which provides
a compact graphical representation of Lyapunov spectra. It also allows
enlightening analogies and differences between spatial and temporal exponents.

CML Lyapunov analysis has several analogies with the
Schr\"odinger problem in disordered lattice systems \cite{localiz,besp}.
This allows extracting from the spatial spectra information about
the localization of the temporal Lyapunov vectors, thus confirming
the close relationship between the two approaches. A discussion of the
analogies with
the Anderson tight-binding problems is presented in section V, where the
the $(\mu,\lambda)$ representation is used for a straightforward
identification of exponentially localized vectors. Moreover, the presence of
of singularities in the spatial Lyapunov spectrum is interpreted as
a stretched-exponential behaviour of the vectors. Finally, another type of
singularity occurring in the temporal spectrum is illustrated. Some general
remarks introducing the reader to the second Part are presented in Section VI.

\section{1D Models for Extended Systems}

Spatio-temporal chaos and instabilities in extended systems
have been widely studied with the aid of simplified models of
reaction-diffusion processes, whose 1D form is of the type \cite{mannev}
\be
\label{cgl}
  \partial_t y = F(y) + D \partial_x^2 y \quad ,
\ee
with the state variable $y(x,t)$ defined on the domain $[0,L]$
(periodic boundary conditions $y(0,t)=y(L,t)$ are generally
assumed). The nonlinear function $F$ accounts for the local reaction dynamics,
while the diffusion constant $D$ represents the strength of the spatial
coupling.

Unfortunately, accurate numerical investigations of partial differential
equations like (\ref{cgl}) can require very long CPU times even on powerful
computers. The introduction of simplified models has been of great help
for understanding the statistical properties of spatio-temporal
chaos.
A first step in simplifying model (\ref{cgl}) is achieved by discretizing
the space variable, i.e. by considering coupled oscillators on a
lattice. This procedure is generally justified whenever the fluctuations
are cut off below a certain spatial scale. After replacing the time
derivative with its
finite difference approximation, one is led to the CML dynamics of the form
\be
\label{readif}
  y^i_{n+1} = F(y^i_n) +(1-\varepsilon)y_n^i
  + {\varepsilon\over 2}\left(y^{i+1}_n+y^{i-1}_n\right) \quad ,
\ee
with $i,n$ being respectively space and time indexes labeling
each variable $y^i_n$ of a lattice of length $L$ and
$\varepsilon=2D$ gauges the diffusion strength.
Model (\ref{readif}) has some technical limitations since, for arbitrary
$F$, it leads to diverging solutions.  To avoid such difficulties one can
postulate a CML dynamics of the form \cite{kaneko,kapral}
\be
\label{mappa}
  y^i_{n+1} = f\left((1-\varepsilon)y^i_n+ {\varepsilon \over 2}
      \left[  y^{i-1}_n+y^{i+1}_n \right]\right)\\ ,
\ee
where $f$ is a nonlinear function mapping a given interval $I$ of the real
axis onto itself. Periodic boundary conditions $y_n^{i+L}=y^i_n$ are
again assumed. A generalization of model (\ref{mappa}) has been
proposed \cite{deissler} to study convective instabilities, namely
\be
\label{amappa}
  y^i_{n+1} = f\left((1-\varepsilon)y^i_n+\varepsilon
      \left[ (1-\alpha) y^{i-1}_n+\alpha y^{i+1}_n \right]\right)\\.
\ee
The parameter $\alpha$ (comprised between 0 and 1) accounts for the
possibility of an asymmetric coupling, mimicking open-flow systems
(first order derivatives in the continuum limit). Obviously, the
symmetric case (\ref{mappa}) is recovered for $\alpha=1/2$.

Coupled maps are sometimes used also as test ground for extended Hamiltonian
systems. The corresponding models are generally formulated in such a way
that the evolution in tangen space is described by symplectic matrices.
As a specific example we will refer to the coupled-standard-maps lattice
\cite{Kantz}
\begin{eqnarray}
   &&p^i_{n+1} = p^i_n - k \left [ \sin q_n^i + {\varepsilon \over 2}
  \sin(q_n^{i+1}-q_n^i) + {\varepsilon \over 2} \sin(q_n^{i-1}-q_n^i) \right]
  \quad ,\nonumber\\
   &&q^i_{n+1} = q^i_n + p^i_{n+1} \quad (\hbox{mod} \quad 2\pi)\quad ,
\label{mapsym}
\end{eqnarray}
where the parameter $k$ rules the amplitude of the nonlinear coupling.

Finally, there is a last class of models that has been used to study evolution
in the presence of a conserved order parameter \cite{cross}. Such CML models
have been introduced in order to make a closer contact with the dynamics of
microscopic conserved dynamical variables in strongly turbulent fluids.
A general scheme for constructing a CML with a conserved quantity is
to impose that $y_{n+1}^i-y_n^i$ is written as a spatial gradient. For
example
\be
\label{mapcon}
  y^i_{n+1} = y^i_n +g(y^{i+1}_n)+g(y^{i-1}_n)-2g(y^i_n)
\ee
preserves the quantity $\sum_i y_n^i$.

\section{Lyapunov Analysis of Spatio-temporal Chaos}

Lyapunov exponents are the main statistical tool for the study of
low-dimensional strange attractors, since they measure the linear instability
of trajectories in phase space. Moreover they can be related to other ergodic
indicators such as entropies and dimensions \cite{eck}. In extended systems,
whenever many interacting modes are present, both temporal and spatial
instabilities must be taken into account. Two complementary approaches have
been developed which are based, respectively, on the temporal (spatial)
growth rate of space (time) periodic perturbations. The former approach is
the standard method to determine the spectrum of Lyapunov exponents. The
latter one, which leads to introduction of spatial exponents, is useful both
in revealing
the localization properties of Lyapunov vectors \cite{localiz} and
characterizing
the stability of open-flow systems \cite{pikov}.

Numerical simulations \cite{limter}  and approximate analytical results
\cite{wayne} indicate that a well defined thermodynamic limit for the
set of Lyapunov exponents is reached for large enough system size, both
in space and time. As a consequence Kaplan-Yorke dimension and
Kolmogorov-Sinai entropy are extensive quantities \cite{grass}, so that the
dynamics of a long chain can be roughly seen as that of many independent
adjacent subchains. This fact permits to define Lyapunov spectra
independent of the system size (see below).

The following two subsections are dedicated to a brief review of the above
mentioned definitions. For simplicity we limit ourself to the symmetric CML
case Eq. (\ref{mappa}), the extension to spatio-temporal flows being
straightforward.

\subsection{Temporal Lyapunov Exponents}

The standard Lyapunov spectrum is obtained by following the evolution of
a perturbation $\delta y_n^i$ in the tangent space with periodic boundary
conditions assumed for it. A more general class of
Lyapunov exponents has been introduced in Ref. \cite{Torcini} by enlarging
the set of allowed perturbations to include exponentially shaped profiles
such as $\delta y_n^i = \Phi_n^i e^{\mu i}$. The evolution equation for
the scaled variable $\Phi_n^i$ reads as
\be
\label{speci}
\Phi_{n+1}^i = m_n^i \left[ {\varepsilon \over 2} e^{-\mu} \Phi_n^{i-1} +
        (1-\varepsilon) \Phi_n^i +
         {\varepsilon \over 2} e^\mu \Phi_n^{i+1} \right]
\quad ,
\ee
where $m^i_n = f'\left( {\varepsilon \over 2}y^{i-1}_n +
(1 - \varepsilon) y^i_n + {\varepsilon \over 2} y^{i+1}_n  \right )$
is the local multiplier, and $\Phi^i_n$ obeys periodic boundary
conditions.
The exponents $\lambda_j(\mu)$ ($1\le j \le L$) obtained from the
iteration of Eq. (\ref{speci}) for a fixed value of $\mu$ will be hereafter
called {\it temporal} Lyapunov exponents \cite{nota}.
The resulting temporal Lyapunov spectrum (TLS) is defined in the
limit $L \to \infty$, as
\be
   \lambda(\mu,n_\lambda)  =  \lambda_j(\mu) \; , \;
   n_\lambda = j/L  \quad
\label{lytemp}
\ee
where the integrated density (the fraction of exponents
smaller than $\lambda$) $n_\lambda$ ranges, by construction, between 0 and 1.
For $\mu = 0$ the standard TLS is recovered.
Exactly the same arguments apply to continuous-time models as, e.g.,
chains of coupled oscillators. When the space too is continuous, the only
difference is that $n_\lambda$ is unbounded from above.

It it well known that the computation of Lyapunov exponents from
Eq.~(\ref{speci}) proceeds through the construction of an
ordered sequence of Lyapunov vectors (i.e. an orthonormal basis), which
generate the most expanding subspaces of increasing dimension \cite{benettin}.
In general, the direction of such vectors depends on the point in phase-space
where they are computed. These fluctuations make an analytic treatment of the
linear stability problem hardly feasible. However, their spatial structure
can be clarified by following in space the evolution of perturbations.
This is precisely the subject of the next section.

\subsection{Spatial Lyapunov Exponents}

Spatial Lyapunov exponents have been introduced as a tool for investigating
the structure of temporal Lyapunov vectors \cite{localiz}. By assuming
a temporal exponential profile for the perturbation,
$\delta y^i_n = e^{\lambda n} \Psi^i_n$, the evolution equation in tangent
space reads as
\be
 \Psi^i_{n+1} = e^{-\lambda} m^i_n \left [{\varepsilon \over 2}  \Psi^{i-1}_n
 + (1- \varepsilon) \Psi^i_n + {\varepsilon \over 2} \Psi ^{i+1}_n
		  \right ] \quad ,
\ee
which, in turn, can be seen as a spatial recursive equation,
$$
\Theta^{i+1}_n = \Psi^i_n
$$
\be
\label{spatial}
\Psi^{i+1}_n = - 2 {(1-\varepsilon) \over \varepsilon} \Psi^i_n +
		 {2e^\lambda \over \varepsilon m^i_n} \Psi^i_{n+1} -
		  \Theta^i_n  \quad ,
\ee
with boundary conditions $\Theta^i_{T+1} = \Theta^i_1$,
$\Psi^i_{T+1} = \Psi^i_1$. Eqs. (\ref{spatial}) is the starting point of a
transfer matrix approach. In fact, by introducing the column vector
${\bf v}(i) \equiv (\Theta^i_1,\Theta^i_2,\ldots,\Theta^i_T,\Psi^i_1,
\ldots,\Psi^i_T)^t$, the spatial evolution can be described in terms of
products of matrices of the form
\be
 J^i = \pmatrix { 0 & 1 \cr -1 & A^i \cr} \quad ,
\ee
where 0 and 1 are the ($T \times T$) null and identity matrix, and $A^i$ is a
bidiagonal matrix, with $A_{n,n}^i  \equiv - 2 (1-\varepsilon) / \varepsilon$
and $A_{n,n+1}^i\equiv 2e^\lambda/ (\varepsilon m^i_n)$.
Since det $J^i$=1 the spatial dynamics is ``conservative" .
Numerical implementation of this method requires the knowledge of the local
multipliers $m^i_n$, which have to be computed, in principle, for a
generic orbit of period $T$. In practice, it is assumed that boundary
conditions are irrelevant in the thermodynamic limit $T \to \infty$, and that
any initial condition chosen according to the invariant measure is almost
equivalent.

For a given $T$, there exist $2T$ Lyapunov exponents $\mu_j$.
This doubling in the number of the degrees of freedom is related
to the two-step spatial memory of Eqs.(\ref{spatial}). Moreover, being
$J^i = (J^i)^{-1}$, the spatial Lyapunov spectrum
(SLS) associated with any symmetric trajectory is
invariant under ``space''-reversal $i \to -i$, i.e. $\mu_j = - \mu_{T+1-j}$.
Notice that this is no longer true for models like (\ref{amappa})
with $\alpha \ne 1/2$.
In analogy to the TLS, the SLS is defined in the limit $T\to \infty$, as
\be
  \mu(\lambda,n_\mu) = \mu_j(\lambda) \; , \;
   n_\mu = {j - 1/2 \over T} - 1 \quad ,
\label{lyspaz}
\ee
where $j$ denotes the $j$-th spatial Lyapunov exponent. The term $1/2T-1$ is a
finite-size correction, introduced to fix the center of symmetry
at $n_\mu = 0$, independently of the size $T$. The density $n_\mu$ of spatial
exponents ranges between -1 and 1.

In analogy with disordered systems, the minimum positive exponent
$\mu(\lambda,0)$ can be interpreted as the inverse of the localization length
$l$ of the Lyapunov vector, provided that $\lambda$ belongs to the TLS.

The direct implementation of the above method faces the difficulty that
iteration of the nonlinear equation in space usually does not converge
onto the same attractor as obtained by iterating in time the original model.
In fact, it turns out that the invariant spatio-temporal measure
corresponds to a strange repeller in space (this point will be
elucidated in the second part). Thus, one must first generate a 2D pattern
of local multipliers and then use it in the computation of the spatial
Lyapunov exponents.

\section{The $(\mu,\lambda)$-space}

A more symmetric representation of both spatial and temporal spectra is
obtained by expressing Eqs. (\ref{lytemp}) and (\ref{lyspaz}) as
$n_\lambda(\mu,\lambda)$ and $n_\mu(\mu,\lambda)$. Both functions are
well defined in so far as $\lambda$ and $\mu$ are monotonously decreasing
for increasing density.
Any point $(\mu,\lambda)$ identifies a specific perturbation growing as
$e^{\lambda n}$ in time and as $e^{\mu i}$ in space. The integrated
density $n_\mu$ ($n_\lambda$) corresponds to the normalized number of
``modes'' with spatial (temporal) exponent smaller than $\mu$ ($\lambda$).

The set $\cal D$ of points in the $(\mu,\lambda)$-plane corresponding to
admissible perturbations is identified by simultaneously requiring.
\be
\label{ineq}
   {\partial n_\mu \over \partial \mu}  \ge 0 \quad , \quad
   {\partial n_\lambda \over \partial \lambda}  \ge 0 \; .
\ee
One should notice a similarity with the dispersion relation for
wave propagation: given a specific spatial profile, the temporal
growth rate is bounded in a given interval from the above inequalities.
The boundaries $\partial {\cal D}_\mu$ and $\partial {\cal D}_\lambda$
of the domain ${\cal D}$ are obtained by setting the l.h.s. equal
to zero in Eq. (\ref{ineq}). In Fig. 1 we report such borders for a
chain of coupled logistic maps ($f(x) = 4x(1-x)$) and $\varepsilon = 1/3$
(a), $\varepsilon = 2/3$ (b). As the two borders coincide in
both cases we have a first indication that $n_\mu$ and $n_\lambda$ are not
independent of one another. Although this is not surprising, since
both quantities arise from the iteration of the same linear set
of equations, it is far from obvious.
The invariance of the dynamical equations under the transformation
$i \to - i$ implies a left-right symmetry of the border $\partial {\cal D}$
(we drop the subscript whenever
$\partial {\cal D}_\mu = \partial {\cal D}_\lambda$).
The upper and lower boundaries at $\mu = 0$ correspond to the standard
maximum $\lambda_{max}$ and minimum $\lambda_{min}$ Lyapunov exponent
(for $\varepsilon=2/3$, $\lambda_{min} = -\infty$ \cite{besp}).
Three typical representatives of temporal and spatial spectra are reported in
Fig.~2 (a-c) and (d-f), respectively, with reference to $\varepsilon=1/3$.
The SLS obtained for $\lambda_{min} \le \lambda \le \lambda_{max}$
is characterized by a single band, while for larger and smaller values
of $\lambda$ two symmetric bands appear.
The TLS appears to be characterized by a single band for any value of $\mu$.
For $\varepsilon=1/3$ the size of the band increases monotonously up to
$\mu = \mu_c$, where the minimum exponent diverges
to $-\infty$, while in the latter case the divergence occurs at
$\mu = 0$. Above $\mu_c$, the size of the band shrinks to zero and its
position increases linearly with $\mu$. \par
This can be easily understood by realizing that, for large $\mu$,
Eq. (\ref{speci}) reads as
\be
\Phi_{n+1}^i \approx m_n^i {\varepsilon \over 2} e^\mu \Phi_n^{i+1}
\quad .
\label{appr}
\ee
Accordingly, the asymptotic value of the Lyapunov exponent is
\be
  \lambda \approx \mu + \log(\varepsilon/2) + \langle \log | m^i_n |
  \rangle \quad ,
\label{lasym}
\ee
where the average $\langle \cdot \rangle$ should be taken along the line
$i = l - n$ for any given $l$. Eq. (\ref{lasym}) is increasingly accurate for
$\mu\to \infty$, when the evolutions along the lines $i=l-n$ are exactly
decoupled, so that an infinite degeneracy in the spectrum is necessarily
found. It is
transparent from Eq. (\ref{lasym}) that the slope of the branch must
be equal to 1. Such a value is nothing but the limit velocity for the
propagation of disturbances in a lattice with nearest-neighbour interactions.
In fact, in the case of a coupling extended to $s$ neighbours,
Eq. (\ref {appr}) becomes, $\Phi_{n+1}^i \approx m_n^i \varepsilon e^{s \mu}
\Phi_n^{i+s}/2$, so that the slope is, in general, equal to the number of
neighbours $s$. Finally, notice that
$\log(\varepsilon/2) + \langle \log | m^i_n | \rangle $ indicates
the intersection of the asymptote with the vertical axis.

The above described phenomenology turns out to be rather general:
simulations performed with several different maps exhibit the
same behaviour as long as one limits the analysis to nearest
neighbour coupling.

An equivalent but instructive representation
is obtained considering $n_\lambda$ and $n_\mu$ as independent
variables. In fact, it turns out that any pair $(n_\mu,n_\lambda)$
identifies unambiguously a spatio-temporal perturbation, exactly in
the same way as $\mu$ and $\lambda$ do.
The two integrated densities are, by construction, normalized in the
rectangle $[-1,+1]\times[0,+1]$. Numerical simulations performed
on logistic maps show that the domain ${\cal D}$ in the
($\mu,\lambda$)-plane corresponds to a domain $\cal B$ which,
strange enough, does not coincide with the full rectangle
and indeed only points with $n_\lambda>|n_\mu|$ are obtained
(see Fig. 3). We have no explanation for this phenomenon.

The four lines identifying the border $\partial {\cal D}$ are mapped
onto the four distinct points $P = (0,0)$, $Q= (0,1)$, $R_+ = (1,1)$
and $R_-= (-1,1)$.  A detailed characterization of the mapping in
presented in Fig. 3. Symmetry reasons require that the
line $\mu = 0$ is mapped onto the line $n_\mu =0$. All vertical lines
in the $(\mu,\lambda)$ plane correspond to curves departing from $P$
in $(n_\mu,n_\lambda)$. Their ending point $E$ depends on $\mu$: for
$\mu<-\mu_c$, $E = R_+$, for $-\mu_c< \mu <\mu_c$, $E = Q$, and for
$\mu>\mu_c$ $E=R_-$. The critical curves for $\mu = \pm \mu_c$ are
straight lines arriving at $(\pm 0.5, + 1)$.

The images of all the horizontal lines connect $R_-$ with
$R_+$ exhibiting an obvious symmetry with respect to $n_\mu$.
All curves corresponding to $\lambda < \lambda_{min}$ pass through
the point $Q$.  Analogously, those ones with $\lambda > \lambda_{max}$
pass through $P$.

We can thus conclude by remarking that there is a sort of duality
between the two representations. In fact, the four segments
delimiting $\cal B$ correspond to 4 points in
the $(\mu,\lambda)$ plane, more precisely to
$(\pm \infty, + \infty)$ and $(\pm \mu_c, -\infty)$ in Fig. 1.

In order to examine the generality of the above scenario
in the following subsections we discuss two specific cases:
(a) spatially homogeneous and stationary pattern, which
can be analytically treated, and (b) spatio-temporal periodic
orbits.

\subsection{Homogeneous Chains}

The simplest chaotic dynamics one can think of is the evolution of
piecewise linear maps of the type $f(x) = rx \quad (\mbox{mod}\; 1)$.
In this case, the local multiplier is everywhere constant (both in space
and time). The resulting Lyapunov spectra coincide with those of
spatio-temporal fixed points (stationary and homogeneous solutions). The
TLS is easily obtained by noticing that the associated eigenvectors
are nothing but the Fourier modes ${\rm e}^{iqj}$ (where $j$ is the imaginary
unit) of the chain.
The TLS can be determined by Fourier transforming Eq.~(\ref{speci}), which
allows calculating the multiplier $m$ for a given wavenumber $q$.
The Lyapunov exponent $\lambda(\mu,n_\lambda) = \log | m |$ is
\be
\lambda(\mu,n_\lambda) = \log r+{1\over 2}\log\left|
(1-\varepsilon)^2+2\varepsilon(1-\varepsilon) \cosh\mu \cos\pi n_\lambda
+\varepsilon ^2 (\cosh^2\mu -\sin^2 \pi n_\lambda) \right| \; ,
\label{spectra1}
\ee
where $n_\lambda$ is equal to the spectral density $q/\pi$, since in virtue of
the node theorem, all the multipliers are naturally ordered with $q$.

The same technique can be used to evaluate the SLS, the only difference
being the two-step spatial memory in Eq.~(\ref{spatial}).
The spectral problem can be solved by determining the eigenvalues of
suitable $2\times 2$ matrices, obtaining
\be
\mu(\lambda,n_\mu) = {1\over 2} \cosh^{-1} \left[
{a +\sqrt{(a-\varepsilon^2)^2 + \varepsilon^2 {\rm e}^{2 \lambda}
   \sin^2(\pi n_\mu)} \over \varepsilon^2} \right ] \quad,
\label{spectra2}
\ee
where
\be
a = {r^2(1-\varepsilon)^2-2r(1-\varepsilon)\cos\pi n_\mu e^{2\lambda}\over 2}
\quad .
\ee
Eqs. (\ref{spectra1},\ref{spectra2}) can be inverted to give the integrated
densities
\be
n^{\pm}_\lambda = {1\over\pi}\arccos \left[{-r(1-\varepsilon)\cosh\mu\pm\sqrt{
r^2(1-2\varepsilon)\sinh^2\mu+e^{2\lambda}}\over \varepsilon r}\right] \; ,
\label{density1}
\ee
\be
n^{\pm}_\mu =
{1\over\pi}\arccos\left[{-r(1-\varepsilon)\sinh^2\mu\pm\cosh\mu\sqrt{
r^2(1-2\varepsilon)\sinh^2\mu+e^{2\lambda}}\over e^\lambda}\right] \quad .
\label{density2}
\ee
For $\varepsilon < 1/2$, only the positive solution exists so that
$n_\lambda = n^+_\lambda$ and $n_\mu = n^+_\mu$. For $\varepsilon >1/2$ both
solutions are acceptable. In fact, the spectrum of multipliers $m$ passes
through zero and then $n(\log|m|)$ is no longer invertible. In this case ,
two branches exist, ordered in an opposite way with respect to $\lambda$.
The integrated density of Lyapunov exponents is obtained by summing the
contributions arising from the two partly overlapping bands,
$$
n_\lambda = n^+_\lambda + (1 - n^-_\lambda)
$$
\noindent
and
$$
n_\mu =  \cases{
n^+_\mu - n^-_\mu \quad &for   $-1 \le n^\pm_\mu \le 0$ \cr
n^+_\mu + (1 - n^-_\mu) &for $0 \le n^\pm_\mu \le 1$}
$$
The analytic expressions of the border $\partial{\cal D}$
can be explicitely obtained by imposing that the argument of the
arccosinus function in $n_\lambda$ is $\pm 1$. For $\varepsilon < 1/2$
this procedure yields
\be
\label{bordo1}
\lambda = \log r +\log\left| \varepsilon\cosh\mu\pm(1-\varepsilon)\right|\quad.
\ee
For $\varepsilon > 1/2$, the upper border is unchanged (plus sign),
while the lower one is now obtained from the identification of the inversion
point where $n^+_\lambda = n^-_\lambda$,
\be
\lambda = \log r +\log\left| \sqrt{(2\varepsilon-1)} \sinh \mu \right|
\quad .
\label{bordo2}
\ee
It should be noticed that the expressions of the borders, as obtained
from Eq. (\ref{density1}) or Eq. (\ref{density2}) are identical.
The two borders reported in Fig. 1 for logistic maps are qualitatively
the same as those obtained in the present case for $\varepsilon <1/2$
and $\varepsilon >1/2$, respectively.

Moreover, from Eqs. (\ref{bordo1},\ref{bordo2}),
$$
 \mu_c =\cases{  \pm \cosh^{-1} \left ( 1 - \varepsilon \over \varepsilon
  \right ) & if $\varepsilon \le 1/2$\cr
  0 & otherwise}
$$
The value of $\mu_c$ is independent of the map $f$. In fact,
the Jacobian matrix can be factorized as the product of two matrices, the
former containing the detail of the local dynamics, the latter being the
diffusion operator, which depends only on $\mu$ and $\varepsilon$. It turns
out that it is precisely the determinant of the latter operator which
vanishes at $\mu = \mu_c(\varepsilon)$. This singular behaviour is associated
with the existence of a Fourier mode, $(-1)^i$, which is killed in just one
iterate.

Extension of the above analysis to asymmetric CML (\ref{amappa}) yields
the following expressions for the borders
\begin{eqnarray}
\label{abordo}
\lambda &= \log r +\log\left| \varepsilon e^{-\mu}+2\alpha\sinh\mu
\pm(1-\varepsilon)\right| \\
\lambda &= \log r + \log \left | \sqrt{\varepsilon^2 - {(1-\varepsilon)^2
\over 4 \alpha (1-\alpha)}}\big(\alpha {\rm e}^\mu - (1 -\alpha)
{\rm e}^{-\mu}  \big) \right |
 \; ,
\end{eqnarray}
which generalize Eqs. (\ref{bordo1},\ref{bordo2}), respectively.
Two examples of boundaries are reported in Fig.~4 for $\alpha = 3/4$
and $\alpha = 1$. The curves are still symmetric,
though now the center of symmetry is
$\tilde \mu = \log[(1-\alpha)/\alpha]/2$.
For decreasing (increasing) values of $\alpha$ from 1/2, the left
singularity moves towards $-\infty$ ($+\infty$) and eventually disappears
at $\alpha = 0$ (1).

Finally, one should notice that all the results contained in this
subsection apply also to the case of a homogenous random evolution:
the only difference is that $\log r$ is to be everywhere replaced
with the average value $\langle \log |m_n^i| \rangle$ .

\subsection{Periodic Orbits}

An interesting representation of the stability properties in the
$(\mu,\lambda)$-plane is obtained for periodic (both in space and time)
orbits. It was already noticed \cite{Torcini} that the standard TLS exhibits
a band structure. This feature generalizes to the presence of
a series of holes within the domain $\cal D$.
The structure of the temporal spectra is
determined from the spatial periodicity and vice versa. More precisely, the
number of bands of the TLS (SLS) can be at most equal to the spatial period
(twice the temporal period).
The borders $\partial {\cal D}$ are
reported in Fig.~5 for a hypothetical orbit of period 3 in space and 2 in time.
Since the reference to a specific CML model is not of particular relevance in
this case, the values of the 6 multipliers have been chosen a priori
without any relation with a preassigned dynamics. As the integrated
densities remain constant inside each hole, the latter ones
are mapped onto single points in the $(n_\lambda, n_\mu)$ plane which are
analogous to $P$, $Q$ and $R_\pm$.

\section{Localization in tangent space}

The localization of Lyapunov vectors observed in Refs. \cite{localiz,livi},
implies that a generic perturbation is sensibly amplified in a limited spatial
region. Heuristically this phenomenon can be explained by noticing
the analogy with the Anderson localization problem. The presence of the
discrete
Laplacian operator makes Eq. (\ref{speci}) similar to a wave equation
on a lattice, or to a Schr\"odinger equation in the tight-binding
approximation.
Chaotic fluctuations of the multipliers play the same role as a time-dependent
disorder in the above problems.\par
A further analogy exists between the spatial evolution and the stationary 2D
Anderson model, defined by the equation
\be
\label{anderson}
\psi_n^{i-1}+\psi_n^{i+1}+\psi^i_{n-1}+\psi^i_{n+1}=(\omega-V_n^i)\psi^i_n
\ee
where $\omega$ is the rescaled energy of the eigenstate $\psi_n^i$ and
$V_n^i$ is the random potential bounded between $-W/2$ and $W/2$.
In this case, both $\omega$ and $W$ can be interpreted as control
parameters in a dynamical equation.
The main difference with Eq. (\ref{spatial}) is that
now also the index $n$ denotes a spatial direction.
The lack of time-reversal invariance in the CML is reflected by
the absence of a term proportional to $\Phi_{n-1}^i$.

On the basis of this simple analogy, one might expect that spatially chaotic
solutions of Eq. (\ref{spatial}), characterized by pseudo-random sequences of
multipliers $m^i_n$, are associated with localized vectors.
In fact, the theoretical analysis of the 2D Anderson model
predicts the existence of exponentially localized eigenfunctions for
arbitrarily weak disorder \cite{abram}.
On the other hand, numerical simulations \cite{localiz} show that
$\mu(\lambda,0)$ is always 0 for $\lambda$ belonging to the standard TLS,
implying that the localization is not purely exponential. This contradiction
can be resolved by noticing that in the CML case, at variance with the
Schr\"odinger problem, the transfer matrices are iterated assuming an
exponential
profile along the transverse direction. Accordingly, different operators
are involved in the two procedures, with an evident symmetry breaking in
the latter case.

Since $\mu(\lambda,0)$ yields the localization length $l$, it is reasonable to
speculate that, whenever $\mu(\lambda,0)=0$, the behaviour of the SLS in the
vicinity of $n_\mu=0$ gives information about a supposedly
lower-than-exponential
decay of the Lyapunov vector. A first necessary condition for this conjecture
to be true is that the convergence (for $T \to \infty$) of the SLS to its
asymptotic shape is sufficiently fast so as to guarantee that the finite-size
value of the minimum positive exponent is essentially
determined by the difference of the corresponding $n_\mu$ from 0, i.e.
\be
  \mu_T(\lambda) \approx \mu \left( \lambda, -{1\over 2T} \right) \; .
\label{finsiz}
\ee
This point will be discussed in the last section of this section; here
we limit ourselves to notice that numerical simulations reveal that the
convergence is not always sufficiently fast.\par
Assuming a power-law behaviour of the SLS, i.e.
\be
\mu \sim |n_\mu|^\gamma \quad n_\mu \to 0 \; ,
\label{gamma}
\ee
and Eq.~(\ref{finsiz}) to hold, then the vector $\delta y^i_n$ decays as
$\exp [ - L/l(T)] \sim \exp [ - L/T^\gamma] $. Whenever there
is a perfect symmetry between spatial and temporal directions, as in
the Anderson model (\ref{anderson}), $T$ can be replaced by $L$ in the
above expression yielding a decay behaviour
\be
  \delta y^L_n \sim \exp[-L^{(1-\gamma)}] \; .
\label{power}
\ee
Accordingly, if $0 < \gamma < 1$, a stretched-exponential envelope
is obtained. We should anyhow recall that $T$ and $L$ are
not in general interchangeable, so that the above picture is only
approximately correct.\par

\subsection{Frozen Random Patterns}

For stationary but spatially chaotic states, the evolution in
tangent space is mapped exactly onto a 1D localization problem.
In this case, it is rigorously proved that all the eigenvectors
are exponentially localized. Thus, qualitative differences
with the scenario depicted in the previous section are expected.
Indeed, frozen random patterns are, to our knowledge, the only
case where the borders derived from spatial and
temporal spectra do not match. This is illustrated in Fig. 6, where
it is clearly seen that the spatial Lyapunov exponents \cite{nota2}
are all bounded
away from zero. As a consequence there is a pair of borders which
cannot be revealed by the temporal analysis (see dashed line in Fig. 6).
In fact, upon increasing $\mu$ from 0, both the maximum and minimum
temporal Lyapunov exponent are constant until the spatial border is
reached (for clarity reasons, only the behaviour of the
maximum exponent is reported in Fig. 6). At larger values of
$\mu$, the two borders
coincide.\par
The anomalous structure of $\cal D$ is a consequence of the exponential
localization of the temporal Lyapunov vectors. In fact, the dashed curve
is nothing but $1/l$ for all temporal exponents belonging to the spectrum.\par
A further interesting feature concerns the identification of the critical
value $\mu_l$, where the upper temporal border departs from the spatial one.
By noticing that the upper border is the Legendre transform
\cite{Torcini} of the maximal comoving Lyapunov exponent \cite{deissler},
it appears natural to conjecture that $\partial {\cal D}_\lambda$ can be
obtained from $\partial {\cal D}_\mu$ by means of a standard construction
to remove all changes of concavity. In other
words we expect $\mu_l$ to be located at the inflexion point of
$\partial {\cal D}_\mu$.
This seems indeed to be true. We shall comment more extensively about this
point in the second part of the paper.

\subsection{Symplectic Maps}

There is a strict analogy between the application of the transfer
matrix approach to the Anderson model (\ref{anderson}) and the computation
of Lyapunov spectra in Hamiltonian systems, since in both cases one deals
with products of ``random'' symplectic matrices \cite{livi}. Generally
speaking, both the symplectic structure
and the time-reversal invariance induce a ``pairing'' in the
Lyapunov spectrum which turns out to be invariant under the transformation
$\lambda \to -\lambda$. Consequently, a similar structure of the
borders is expected along the $\mu$ and $\lambda$ directions.
Obviously, in the Anderson model both are spatial directions
and they can be exactly interchanged, so that we expect the exact
invariance under the transformation $(\mu,\lambda) \to (\lambda,\mu)$. \par
The main difference with respect to the previous class of dissipative CML
is the slope of the lower asymptotes, which are now tilted at a
$\pm\pi/4$ angle.
This is precisely the consequence of the invertibility of mapping
(\ref{mapsym}). Indeed, when the evolution is reversed in time, the maximum
velocity of propagation of a disturbance is 1 (when
nearest-neighbour coupling is assumed). This fact allows to interpret
the vertical asymptotes in Fig.~1 as an indication of an infinite (backward)
velocity. The non-invertibility of the dynamics leads to a global
unavoidable ambiguity of the preimage already in one step.\par
Simulations of model (\ref{mapsym}) show two qualitatively different
domain structures. Indeed, depending on the nonlinearity $k$,
a hole may exist around $(\lambda,\mu) = (0,0)$, indicating
that the standard TLS is made of two distinct bands.
The two possibilities are illustrated in Fig.~7, where $\partial{\cal D}$
is reported
for $k=0.5$ and $k=4.0$. At variance with the latter case, at $k=0.5$
a substantial disagreement is found between spatial and temporal borders.
This artifact seems to be caused by strong finite-size effects in the
convergence of the SLS to its asymptotic form. This may be connected with
the slowly decaying temporal correlations (up to time of order $10^3$)
previously revealed by the analysis of diffusion properties \cite{kancon}
and of effective Lyapunov exponents \cite{Kantz}.

To close this section, we want to remark that the theory developed in
\cite{abram} for the 2D-Anderson problem, predicting that all states
are exponentially localized, would imply the presence of a hole
at $(\lambda,\mu) = (0,0)$ for any value of $k$ in the present case. On the
other hand, direct numerical simulations of model
(\ref{anderson}) appear to be quite inconclusive due to the rapid
divergence of the localization length for $W \to 0$ \cite{pichard}.
Thus, also in our dynamical model it is not possible to conclude whether there
are two truly distinct regimes as suggested by Fig.~7.

\subsection{Localization and Spatiotemporal Chaos}

In this subsection we turn again to the general case of
spatio-temporal chaos, reporting some results on the power-law
singularity in the SLS. In general, for $\lambda$ belonging to the
standard TLS, the corresponding SLS exhibits a linear
behaviour around $n_\mu = 0$, that is $\gamma = 1$ (see
Eq.~(\ref{gamma})). Hence, according to Eq.~(\ref{power}), the Lyapunov vector
decays more slowly than a stretched exponential. In fact, direct numerical
investigations reveal a power-law localization \cite{localiz2}. \par
The only exception to the above picture is represented by the extrema of
the standard spectrum, i.e. $\lambda = \lambda_{min}$, $\lambda_{max}$,
where $\gamma$-values substantially smaller than 1 are observed.
This is not surprising in that these are the points where a gap is
closing in the SLS (see for example Fig. 1).  Although this
singularity arises at a band merging point in the $(\mu,\lambda)$ plane,
$\gamma$ is not trivially linked with the shape of the boundary around
$\mu=0$. In fact, while $\gamma$ may depend on the model, the shape of
$\partial {\cal D}$ around $\mu = 0$ turns out to be always parabolic.
This has been verified both analytically for homogeneous chains by
expanding Eq.~(\ref{spectra2}) around $\mu = 0$, and numerically for
various models (see Fig.~8, where
$\delta \lambda = \lambda(\mu,0) - \lambda_{min}$ is plotted versus $\mu$).\par
Extensive numerical simulations  have been performed on
several classes of CMLs to measure $\gamma$, namely model (\ref{mappa}) with
logistic, tent, skewed and Bernoulli maps, model with linear coupling
(\ref{readif}), and symplectic maps (\ref{mapsym}). The mapping functions as
well as
the measured values of $\gamma$ are reported in Table 1,
indicating that the exponent is either approximately equal to 1/3 or to 1/2.
The inaccuracy is essentially to be attributed to finite-size effects, as
the maximum temporal exponent has been always computed with a relative
error of $10^{-3}$. Actually, the above estimates can be improved by
assuming that the spectrum is an analytic function of $n_\mu^\gamma$.
More precisely, the power-series expansion of $\mu$ to the second order,
\be
\mu\approx a n_\mu^\gamma + b n_\mu^{2\gamma} \quad ,
\ee
can be used to fit the numerical data in the vicinity of $\mu = 0$,
in order to determine the parameters $a$ and $b$ for a given $\gamma$.
The best estimate of the exponent $\gamma$ is finally determined
as the one which optimizes the fit. As a result of this application
the $\gamma$-values are even closer to 1/3 and 1/2 confirming that
there are two universality classes. The analysis of the considered models
reveals that the exponent 1/2 is obtained only when the local variable is
scalar and if the local multipliers are all positive (negative). Hence
the exponent 1/3 seems to arise from ``dephasing'' effects due to a local
rotation of the perturbation (if the variable is a vector) or to the
randomness of the sign of the multipliers (if the variable is a scalar).
The larger generality of the latter value is further confirmed by its
observation in very different contexts such a band random matrices \cite{tsamp}
or nonequilibrium molecular dynamics \cite{posch}.\par
A further interesting interpretation of the two universality classes appears
from the structure of the temporal evolution in tangent space
(Eq.~(\ref{speci})). Pikovsky and Kurths \cite{pikovpre} have shown that
the equation resulting from the change of variable $h^i_n = \ln \delta y_n^i$
is approximately equal to a discrete version of the Kardar-Parisi-Zhang
equation \cite{kpz} for the scalar field $h(x,t)$ ($x$ and $t$ replace $i$,
$n$, respectively)
\be
   \partial_t h = \eta_1 \partial_{x}^2 h + \eta _2
   \big( \partial_{x} h \big)^2 + \xi
   \; ,
\label{kpz}
\ee
where $\xi(x,t) = \ln m^i_n$ is a noise term arising from space-time chaos,
and $\eta _{1,2}$ are two constants.
A first limitation to an exact correspondence stems from higher order spatial
derivatives the presence of which should not affect the scaling behaviour
\cite{kpz}. A more relevant difference arises when the sign of $m_n^i$ is
a fluctuating quantity, since the definition itself of $h_n^i$ is meaningless,
as already argued in \cite{pikovpre}. According to our previous
considerations, $\gamma$ turns out to be 1/2 if and only if the reduction to
Eq.~(\ref{kpz}) is possible. In such cases, from Eq.~(\ref{power}), we expect
for the Lyapunov vector a stretched exponential profile
$\delta y_n^L \sim \exp (- L^{1/2})$, which is perfectly consistent with the
prediction that the field $h$ describes a Brownian motion in space.
Thus, the presence of the exponent $\gamma = 1/3$ seems to call for a
different stochastic model.\par
Finally, let us recall that the exponent $\gamma$ can provide information about
the localization of $\delta y_n^i$ only when the convergence of the SLS is
sufficiently fast. For instance, the exponent $\gamma = 1/2$ is observed
also for a homogeneous chain, where no localization is obviously present.
The apparent contradiction is solved by noticing that in this case
the $T$-th spatial exponent is exactly equal to 0 for any $T$. Therefore,
$\gamma$ does not describe the convergence of the minimum positive exponent
as expected, and the above arguments are no longer valid.

\subsection{Singularities in the TLS}

For the sake of completeness, we briefly discuss also an example of
a singularity occurring in the TLS, which is not related to localization
features of Lyapunov vectors, but rather to the existence of a conserved
parameter.\par
All singularities in SLS discussed in the previous section are
characterized by $\gamma <1 $, that is by a vanishing density of
Lyapunov exponents. The study of TLS in models with a conserved
quantity, such as Eq.~(\ref{mapcon}), has revealed the existence of
the complementary phenomenon too, i.e. a power-law divergence in the
density of exponents
\be
{ dn_\lambda \over d\lambda} \sim {1 \over |\lambda |^\beta}  \; ,
\ee
around $\lambda=0$ \cite{bohrgrin}. By calling $n_c$ the integrated
density determined by the condition $\lambda(0,n_c) = 0$, the above
equation can be rephrased as $|\lambda| \sim |n_\lambda - n_c|^\nu$ with
$\nu = 1/(1-\beta)$. Fig.~10 shows the scaling behaviour around
the critical integrated density $n_c$ for model (\ref{mapcon})
with $g(x)= \varepsilon_1 \sin(2\pi x)+ \varepsilon_2 x$  ($\varepsilon_{1,2}$
are two coupling constants). Crosses and diamonds correspond to approaching
$\lambda = 0$ from above and below, respectively. The different slopes
($\nu$ about 1.5 in the former case and 2 in the latter) confirm the
existence of two distinct critical behaviours as conjectured in
\cite{bohrgrin}. This phenomenon appears to be very general, although a
global explanation is still lacking.\par

\section{Concluding remarks}
In this first part, we have introduced and utilized a general representation
for describing the linear evolution of perturbations in spatially extended
systems. We claim that spatial and temporal spectra allow for a comprehensive
description of all phenomena occurring in this framework. In the second
part we shall show how the comoving and rotated exponents mentioned in the
Introduction can be derived from SLS and TLS. At the
present moment it is still unclear to what extent spatial and temporal
exponents
are independent of one another. As far as the border $\partial {\cal D}$ in the
$(\mu,\lambda)$-plane is concerned, the two classes are practically equivalent.
However, no relation has yet been found between the corresponding densities.

\acknowledgments

We thank Giovanni Giacomelli, Peter Grassberger, Holger Kantz,
and Arkady Pikovsky for useful discussions.
One of us (A.T.) gratefully acknowledges the European Economic Community
for the research fellowship No ERBCHBICT941569 ``Multifractal Analysis
of Spatio-Temporal Chaos''.

\newpage



\begin{figure}
FIG. 1: Plot of the boundary $\partial{\cal D}$ for the logistic
CML for two values of the coupling (a) $\varepsilon=1/3$,
(b) $\varepsilon=2/3$.
\end{figure}

\begin{figure}
FIG. 2: (a-c) Temporal Lyapunov spectra for the logistic CML
($\varepsilon=1/3$) for $\mu = 0$ , 1.31 (i.e. $\mu_c$) and 3.0,
 respectively;
(d-f) Spatial Lyapunov spectra for the same diffusive coupling and
$\lambda = -2.0$ , 0 and 2.0, respectively.
\end{figure}

\begin{figure}
FIG. 3: Schematic plot of the contour lines of $\lambda$ (solid line)
and $\mu$ (dashed line) and the boundary $\partial{\cal B}$ in the
$(n_\mu,n_\lambda)$-plane .
\end{figure}

\begin{figure}
FIG.4: Plot of the boundary $\partial{\cal D}$ for the asymmetric
Bernoulli chain for $\varepsilon=1/3$ : (a) $\alpha=3/4$,
(b) $\alpha=1$.
\end{figure}

\begin{figure}
FIG. 5: Plot of the boundary $\partial{\cal D}$ for a spatiotemporal
periodic orbit (period 3 in space and 2 in time).
\end{figure}

\begin{figure}
FIG. 6: Plot of the boundary $\partial{\cal D}$ for a frozen random
pattern. Dots and lines refer to temporal and spatial exponents
respectively. Below $\mu = \mu_l$,
$\partial{\cal D}_\lambda \ne \partial{\cal D}_\mu$.
\end{figure}

\begin{figure}
FIG. 7: Plot of the boundary $\partial{\cal D}$ for coupled
standard maps (a) $k=0.5$ (b) $k=4.0$. Symbols refer to spatial
exponents while the line is obtained from the temporal spectrum.
\end{figure}

\begin{figure}
FIG. 8: Scaling of the maximum temporal exponent around $\mu=0$,
logistic maps with random (circles) or deterministic (diamonds) dynamics,
$\varepsilon=1/3$. The dashed line has slope 2.
\end{figure}

\begin{figure}
FIG. 9: Singularity of the spatial spectra ($\varepsilon=1/3$):
(a) skewed piecewise CML with deterministic (diamonds)
and random (crosses) multipliers, cubic CML (squares)
(1/2 singularity);
(b) logistic CML with deterministic (triangles) and random
(plus) multipliers, symplectic CML (circles)
(1/3 singularity). Details on the mapping functions and the
best-fit values of the exponents are reported in Table 1.
\end{figure}

\begin{figure}
FIG. 10: Singularity of the temporal spectra of CML
(\ref{mapcon}) with conserved quantity:
$g(x)= \varepsilon_1 \sin(2\pi x)+ \varepsilon_2 x$,
$\varepsilon_1=1/5$ and $\varepsilon_2=1/20$.
The critical value of the integrated density is
$n_c =0.643$.
\end{figure}

\begin{table}
\vskip 1 truecm
\begin{tabular}{lc}
\multicolumn{1}{l}{CML model}
& \multicolumn{1}{c}{$\gamma$}  \\
\hline
\hline
 \\
 tent             & 0.37  \\
 tent (rnd)       & 0.36  \\
 logistic         & 0.32  \\
 logistic (rnd)   & 0.32  \\
 linearly coupled & 0.32  \\
 symplectic       & 0.31  \\
\\
\hline
\\
 skewed Bernoulli & 0.54 \\
 skewed Bernoulli (rnd)& 0.45\\
 cubic            & 0.49  \\
\\
\end{tabular}

\vskip 1 truecm

\caption[tabone]{
Singularity exponent $\gamma$ of the SLS for various
CML models Eq.~(\ref{mappa}) with the mapping function
of the interval $[0,1]$ being
$f(x) = 1-|2x-1|$ (tent),
$f(x) = 4x(1-x)$ (logistic),
$f(x) = ~3x$ for $x<1/3$ and $f(x) = (3x-1)/2$ otherwise
(skewed Bernoulli),
$f(x) = 3/2 x +x-x^3$ (mod~ 1) (cubic).
Linearly coupled maps refer to Eq.~(\ref{readif}) with
$F(x) = 4x/(x^4+1)$ and symplectic to Eq.~(\ref{mapsym})
with $k=4.0$. In every case a coupling $\varepsilon=1/3$ is taken.
The label ``rnd'' refers to random multipliers,
with the same distribution as that of the corresponding CML.
}
\label{tab1}
\end{table}


\begin{references}
\bibitem{nota} In Ref. \cite{Torcini} they were named ``specific''
exponents.
\end{references}
\end{document}